\documentclass[10pt,prc,aps]{revtex4}
\UseRawInputEncoding
\draft
\usepackage{graphicx}
\usepackage{color}
\usepackage{mathtools}
\begin{document}

\title{Neutron skin systematics from microscopic equations of state  }       
\author{            
Francesca Sammarruca }                                                         
\affiliation{ Physics Department, University of Idaho, Moscow, ID 83844-0903, U.S.A. 
}

\begin{abstract}

This paper presents an analysis of neutron skins based on fully microscopic equations of state, including chiral two- and three-nucleon forces consistently at N$^3$LO. Other theoretical predictions and recent constraints are also addressed, such as those from the PREX II experiment and the latest parity-violating electron scattering measurement of the $^{27}$Al neutron skin.  

\end{abstract}
%\keyword{}
\maketitle

\section{Introduction} 
\label{Intro} 

Recently~\cite{SM22}, neutron star predictions based on the most recent neutron and nuclear matter equations of state (EoS) developed by myself and collaborators were reported. Extensive descriptions of the EoSs, which include all subleading chiral three-nucleon forces~\cite{SM21,SM_2_21}, and their application to medium-mass neutron stars can be found in those references and will not be repeated. 
It suffices to briefly recall that 
chiral effective field theory (EFT)~\cite{Wei92,Wei90} provides a path to a consistent development of nuclear forces. Symmetries relevant to low-energy QCD, in particular chiral symmetry, are incorporated in the theory. Thus, although the degrees of freedom are pions and nucleons instead of quarks and gluons, there exists a solid connection with the fundamental theory of strong interactions through its symmetries. The predictions for the symmetry energy (and related properties) utilized here are fully microscopic and employ high-quality, non-local two-nucleon forces (2NF) up to fourth order of the chiral expansion~\cite{EMN17} together with all chiral three-nucleon forces (3NF) required at each order. See Refs.~\cite{SM21,SM_2_21} and citations therein for details of how 3NFs are implemented. Here, the focus is, instead, on the neutron skins of several neutron-rich nuclei implied by the microscopic EoS using a droplet model that includes surface width contributions.

The formation of the neutron skin, 
\begin{equation}
S = <r^2>_n^{1/2} -  <r^2>_p^{1/2} \; ,
\end{equation}
is a fascinating phenomenon. It is the result of excess neutrons being pushed outward by the neutron-enriched core of the nucleus, which effectively generates a pressure gradient. Although a small contribution to the size of the nuclear radius, the neutron skin reveals important information about the physics of nucleon interactions with changing density -- hence, its close relationship to the EoS of isospin-asymmetric nuclear matter and the structure of neutron stars, which is determined by the EoS of neutron-rich $\beta$-stable matter.

Relating the nucleus' spatial extension as directly as possible to the microscopic EoSs is best achieved by means of the droplet model. This also allows applications to heavy nuclei, which may be outside the reach of {\it ab initio} methods.

The purpose of this paper is to:
\begin{enumerate}
\item
Calculate the neutron skin of $^{208}$Pb using droplet model expressions where the symmetry energy and its density slope at saturation appear explicitely. 
\item 
Compare with other predictions and recent constraints, such as those extracted from the
 PREX II experiment~\cite{prexII} for $^{208}$Pb and Ref.~\cite{27Al} for $^{27}$Al.
\item
Consider other nuclei to explore the isospin-asymmetry dependence of the skin and of the surface width contribution. 
\item 
After years of searching for the density dependence of the symmetry energy, large experimental effort addressing the question through diverse methods, and a multitude of phenomenological (Skyrme or relativistic mean field) models generating correlations among the relevant quantities (see Refs.~\cite{rev1}-\cite{rev13} for some examples), I would like to reflect on the best way forward given modern predictions and recent constraints, especially those extracted from electroweak scattering experiments.
\end{enumerate}

The paper consists of the following sections. Section~\ref{ldm} contains a brief review of the droplet model utilized in this work. Results are presented and discussed in Section~\ref{res} along with other theoretical predictions and empirical constraints. A  summary and conclusions are found in Section~\ref{Concl}.

\section{The neutron skin in the droplet model} 
\label{ldm} 

In the traditional version of the droplet model~\cite{DM1,DM2,DM3}, the neutron skin is written as 
\begin{equation}
S_0 = \sqrt{\frac{3}{5}} \Big (t - \frac{e^2Z}{70\; J} \Big ) \; ,
\label{S1}
\end{equation}
where $e$ and $Z$ are the electron charge and the nucleus' proton number, respectively;
 $t$ is a measure of the separation between the mean locations of the neutron and proton surfaces, and $J$ is the symmetry energy at saturation density. 
  $t$ has the form:
\begin{equation}
t= \frac{3}{2}r_0 \frac{J}{Q} 
\frac{I - cZ/(12JA^{1/3})}
{1 + 9J/(4QA^{1/3})} \; ,
\label{S2}
\end{equation}
with $I$ is the isospin asymmetry of a nucleus with $N$ neutrons and mass number $A$, $I=(N-Z)/A$,  $Q$ the surface stiffness coefficient, and
 $c = \frac{3e^2}{5r_0}$, where $r_0$ is the proportionality coefficient between $A^{1/3}$  and the nuclear radius, approximately 1 fm. 

Equation~(\ref{S1}) does not include the surface width contribution, which was found to be sizable~\cite{Warda2009}. It is written as~\cite{DM1,DM2,DM3}:
\begin{equation}
\Delta S= \frac{5}{2R}(b_n^2 - b_p^2) \; , 
\label{delS}
\end{equation}
which clearly vanishes if $b_n$ and $b_p$ -- the surface widths of the neutron and protons density distributions --  are equal, and becomes larger with growing isospin asymmetry. Clearly, $b_n$ = $b_p$  amounts to the assumption that the proton and neutron surfaces are shifted relative to each other but have the same thickness. Including this mechanism, the skin is
\begin{equation}
S = S_0 + \Delta S \; .
\label{S3}
\end{equation}

One can see from Eq.~(\ref{S2}) that, for large $A$, $t$ is approximately linear with $J/Q$, suggesting a correlation between the skin and $J/Q$ in heavy nuclei~\cite{Warda2009}, to leading order. Furthermore, a correlation between $\Delta S$ and $x=J/Q$ was also determined, which lies between~\cite{Warda2009} 
\begin{equation}
\Delta S = [(0.3x -0.05)I]\; \; {\rm fm} \; \; \; \mbox{and}\; \; \;  \Delta S = [(0.3x +0.07)I]\; \; {\rm fm}  \; . 
\label{S4}
\end{equation}
Thus, to leading order, both $t$ and $\Delta S$ are driven by $x$, which is then a natural parameter to explore model dependence. At the same time, a linear regression between $x$ and the slope parameter $L$ (in MeV) was found to lie between~\cite{Warda2009}
\begin{equation}
L = (139x -52)\; {\rm  MeV} \; \; \; \mbox{and} \; \; \;   L = (150x -57) \; {\rm MeV}  \; . 
\label{S5}
\end{equation}
Using Eq.~(\ref{S5}) to express $x$ in terms of $L$, the neutron skin will then be a function of $J$ and $L$, two crucial parameters in the expansion of the symmetry energy around the saturation density, $\rho_0$, with respect to $\delta = \frac{\rho -\rho_0}{\rho_0}$,
\begin{equation}
e_{sym}(\delta) \approx J + \frac{L}{3}\delta + ... 
\label{S6}
\end{equation}

\section{Results and Discussion }
\label{res}

\subsection{Neutron skin of $^{208}$Pb: predictions and constraints}
\label{Pb}

Using the equations displayed in the previous section, one can predict the neutron skin for given $J$ and $L$. Accounting for the range of parameters given in Eqs.~(\ref{S4}-\ref{S5}), I determine, for the specified approach, a range of values for the neutron skin of $^{208}$Pb. These are shown in Table~\ref{tab1}. For the calculations based on Ref.~\cite{SM22}, the values of $Q$, the surface stiffness coefficient, fall between 39.0 MeV and 45.6 MeV.

\setlength{\tabcolsep}{18pt}

\begin{table}[t!]
\caption{ The neutron skin of $^{208}$Pb, $S$, calculated as described in Sec.~\ref{ldm} using the specified symmetry energy, $J$, and its slope at saturation, $L$. 
}
\label{tab1}
%\begin{tabular*}{\textwidth}{@{\extracolsep{\fill}}cccc}
\begin{tabular}{c c c c }
\hline
\hline
  $J$ (MeV)        & $L$ (MeV)  & $S$ (fm)  & source for $J$, $L$   \\
\hline    
\hline 
 31.3 $\pm$ 0.8  &  52.6 $\pm$ 4.0  & 0.13...0.17 & \cite{SM22}   \\
31.1...32.5   &  44.8...56.2  & 0.12...0.17 & \cite{drisch19}   \\
28...35  &  20...72  & 0.078...0.20 & \cite{DHW21}    \\
27...43                & 7.17...135  &  0.055...0.28           &  \cite{LH19}    \\     
 38.29  $\pm$ 4.66  &  109.56  $\pm$ 36.41  &  0.17...0.31 &  \cite{prexII}   \\              
\hline
\hline
\end{tabular}
\end{table}

\begin{table}[t!]
\caption{ The isospin asymmetry, $I=(N-Z)/A$, for several nuclei, predictions of the neutron skin, $S$, and empirical values obtained from a linear fit to the data given in Ref.~\cite{DM2}. 
}
\label{tab2}
\begin{tabular}{c c c c}
%\begin{tabular*}{\textwidth}{@{\extracolsep{\fill}}cccc}
\hline
\hline
 Nucleus        & I  & $S$ (fm)  &  $S$ from fit to data~\cite{DM2}  \\
\hline    
\hline 
 $^{58}$Ni  &  0.034 & 0.0044...0.011 & -0.025...0.026 \\
$^{27}$Al  &  0.037 & 0.016...0.023 & -0.022...0.029  \\
$^{59}$Co  &  0.085 & 0.046...0.063 & 0.014...0.079   \\
$^{90}$Zr  &  0.11  & 0.061...0.084 &  0.033...0.106 \\
$^{48}$Ca  &  0.17  & 0.12...0.15  &  0.075...0.165 \\
\hline
\hline
\end{tabular}
\end{table}

\begin{table}[t!]
\caption{ Isospin asymmetry and predicted neutron skin, $S$, of Tin isotopes. The data are from Ref.~\cite{Tera08}.
}
\label{tab3}
\begin{tabular}{c c c c}
%\begin{tabular*}{\textwidth}{@{\extracolsep{\fill}}cccc}
\hline
\hline
  Nucleus        & I  & S (fm)  &    Data  \\
\hline    
\hline 
 $^{116}$Sn  &  0.14 & 0.079...0.11 &  0.110 $\pm$ 0.018 \\
$^{118}$Sn  &  0.15 & 0.091...0.12 &  0.145 $\pm$ 0.016 \\
$^{120}$Sn  &  0.17  & 0.10...0.14 &  0.147 $\pm$  0.033 \\
$^{122}$Sn  &  0.18  & 0.12...0.15 &   0.146 $\pm$ 0.016 \\
$^{124}$Sn  &  0.19  & 0.13...0.17  & 0.185 $\pm$ 0.017 \\
\hline
\hline
\end{tabular}
\end{table}

The first three entries in Table~\ref{tab1} are obtained from EoS based on chiral EFT, with chiral two- and three-nucleon interactions at N$^3$LO. One can see that they are relatively soft, cover a narrow range, and are in good agreement with one another. The fourth line correspond to an analysis based on current constraints from nuclear theory and experiment. 
In Ref.~\cite{LH19} (next line in the Table), the authors utilized 48 phenomenological models, both relativistic mean field and Skyrme Hartree-Fock. 
The last line shows the values of $J$ and $L$ from the recent PREX II experiment. Note that the reported value for the skin of $^{208}$Pb  in Ref.~\cite{prexII} is (0.283 $\pm$ 0.071) fm.

A range for $L$ between 45 MeV and 65 MeV is typical (and perhaps somewhat generous) for state-of-the-art nuclear theory, with no overlap with the PREX II result. The corresponding neutron skins are then relatively small. Most recently, {\it ab initio} predictions for the neutron skin of $^{208}$Pb have become available~\cite{Hu+21}. The reported range is between 0.14 fm and 0.20 fm -- smaller than the values extracted from parity-violating electron scattering. On the other hand, a much larger range for the neutron skins can be obtained with phenomenological interactions, both relativistic and non-relativistic mean-field models, including values that are consistent with the PREX II findings. This is to be expected, because much larger variations in $L$ are allowed by mean-field models. Some may argue that the realistic nature of few-nucleon forces doesn't need to be preserved in heavier systems -- hence, the large variations in the properties of the EoS, which  are unconstrained by low-energy nucleon-nucleon data. This argument is incorrect, both on principle grounds -- based on the {\it ab initio} philosophy -- and in practice, as demonstrated in Ref.~\cite{Hu+21}. For these reasons, mean-field models, while  remaining an important tool to explore sensitivities and correlations, lack the predictive power needed to shed light on open questions in {\it ab initio} nuclear structure.

 Neutron skins are calculated in Ref.~\cite{CC} with coupled-cluster theory using the interaction from Ref.~\cite{n2losat} and the ones from Ref.~\cite{GO}, and with the auxiliary field diffusion Monte Carlo method for 
 the local interaction at N$^2$LO from Ref.~\cite{QMC}. With those interactions, they predict a range for $L$ between 58.4 MeV and 65.2 Mev, and values for the neutron skin of $^{48}$Ca between 0.114 and 0.186 fm. To gauge this method, I calculated the neutron skin for $^{48}$Ca using values of $J$ and $L$ from $\Delta$N$^2$LO$_{GO}$ and obtained a range of $S$ between 0.13 fm and 0.16 fm (including both values of the cutoff, $\Lambda$= 450 and 394 MeV), which is within the larger interval given in Ref.~\cite{CC}. As a further verification of the method, I used the values of $J$ and $L$ from Ref.~\cite{CC2016}, $(25.2 \leq J \leq 30.4)$ MeV and $(37.8 \leq L \leq 47.7)$ MeV, and found the range $(0.11 \leq S \leq 0.14)$ fm, to be compared with $(0.12 \leq S \leq 0.15)$ fm from Ref.~\cite{CC2016}.

Before leaving this section, it may be useful to recall that the radius of a neutron star with M=1.4 ${\rm M}_{\odot}$ comes out between 11 and 13 km~\cite{SM22} with the EoSs applied here.

\begin{figure*}[t!]
\centering
\hspace*{-1cm}
\includegraphics[width=7.5cm]{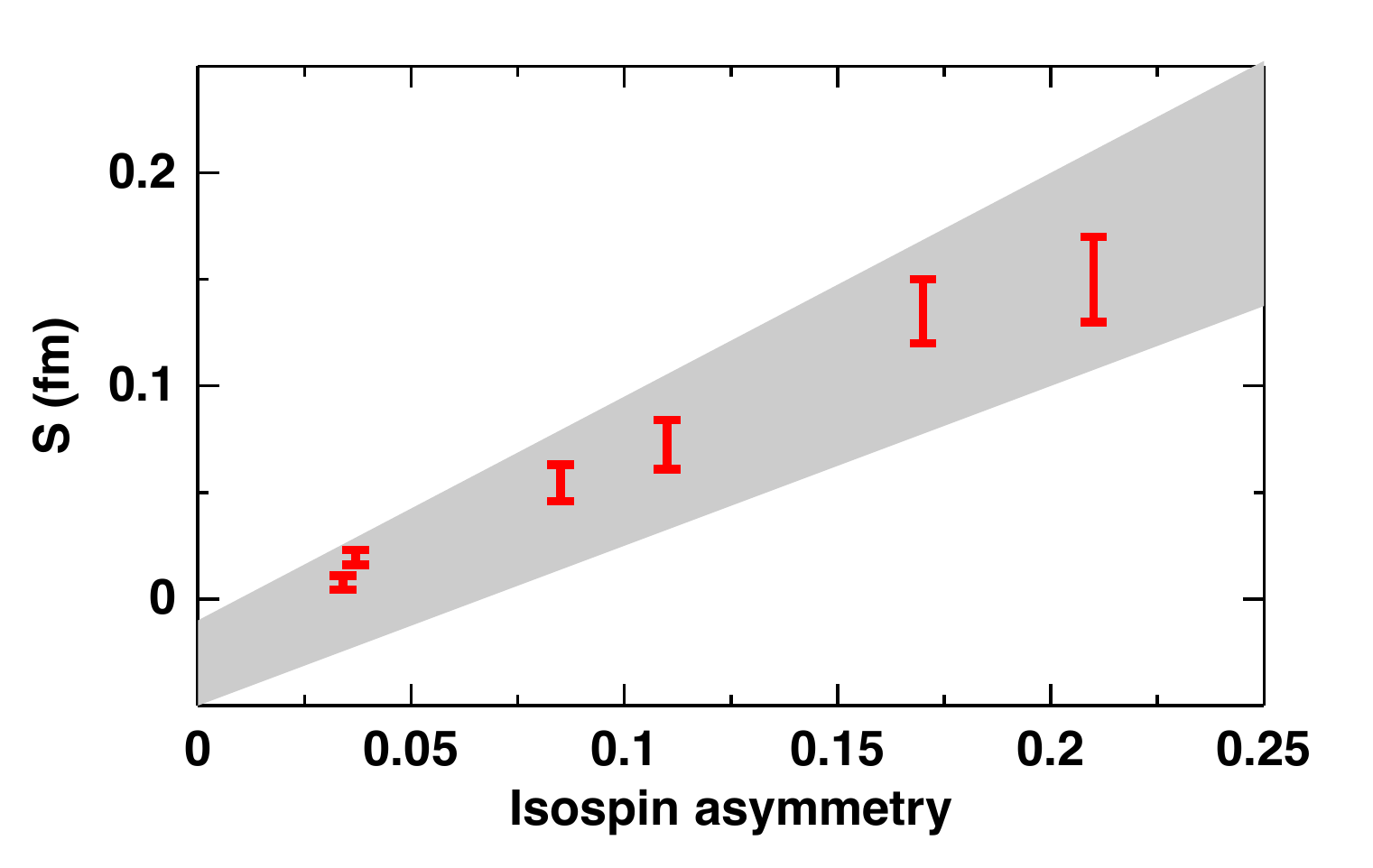}\hspace{0.01in}
\vspace*{0.05cm}
\caption{ Red bars: Neutron skin of  $^{58}$Ni, $^{27}$Al, $^{59}$Co, $^{90}$Zr,  $^{48}$Ca, and 
$^{208}$Pb, in order of increasing isospin asymmetry, as predicted in Refs.~\cite{SM21, SM_2_21}. The shaded area is bounded by linear fits to the data~\cite{DM2}. 
}
\label{S_iso}
\end{figure*}

\subsection{Other neutron-rich nuclei}
\label{other}

In this section,  a variety of neutron-rich nuclei are considered for the purpose to observe the pattern of the neutron skin with growing neutron excess, see Table~\ref{tab2} and Fig.~\ref{S_iso}.
In Fig.~\ref{S_iso}, the shaded area is bounded by linear fits to the data~\cite{DM2}, while the red bars are the predictions obtained in this work for the nuclei in Table~\ref{tab2} in order of increasing isospin asymmetry, $^{58}$Ni, $^{27}$Al,  $^{59}$Co, $^{90}$Zr,  $^{48}$Ca, and 
$^{208}$Pb.

\begin{figure*}[t]
\centering
\hspace*{-1cm}
\includegraphics[width=7.5cm]{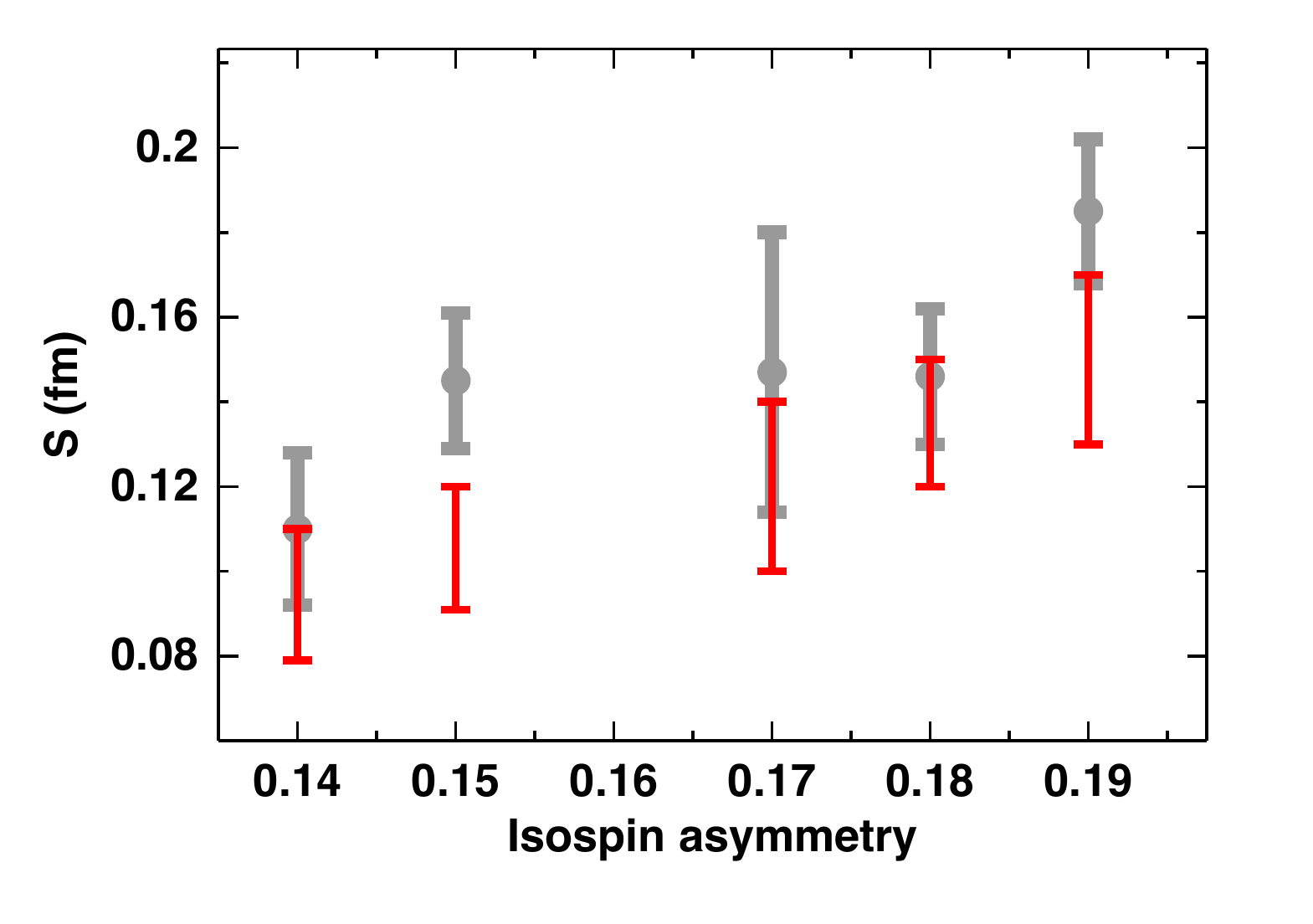}\hspace{0.01in}
\vspace*{0.05cm}
\caption{ Red bars: neutron skins of Tin isotopes, $^{116}$Sn, $^{118}$Sn, $^{120}$Sn, $^{122}$Sn,
and $^{124}$Sn, as predicted with the EoS of Refs.~\cite{SM21, SM_2_21}. Gray bars: data from Ref.~\cite{Tera08}.
}
\label{Tin}
\end{figure*}

Isotopes of Tin are shown separately, see Table~\ref{tab3} and Fig.~\ref{Tin}, to better capture the evolution of the skin across this remarkable chain, which contains the largest number of stable isotopes.The data points were extracted from proton elastic scattering on Tin isotopes at 295 MeV~\cite{Tera08}.

The neutron skin of $^{27}$Al was recently extracted from parity-violating electron scattering off this nucleus. The reported result is $S$ =( -0.04 $\pm$ 0.12) fm, or $S$ between -0.16 and +0.08 fm, which is consistent with a value of nearly zero but also allows for large negative values and correspondingly large positive values for the proton skin.This seems unlikely. Note that, even though this isotope has only one extra neutron above the number of protons, its isospin asymmetry, $I$, is slightly larger than Nickel's. Thus, the predicted skin fits smoothly within the nuclei shown in Table~\ref{tab2} and on the approximately linear behavior apparent from Fig.~\ref{S_iso}, as to be expected from the liquid droplet model, where quantum effects are averaged out. 

In closing this section, I show the isospin asymmetry dependence of the surface width contribution, see Table~\ref{tab4}. As can be expected, $\Delta S$ becomes a more significant fraction of the total skin with increasing neutron excess -- over 30\% around the  region of Lead.

\begin{table}[t!]
%\setlength{\tabcolsep}{10pt}
%\renewcommand{\arraystretch}{1.5}
%\begin{tabular*}{\columnwidth}{@{\extracolsep{\fill}}cc}
\caption{ Isospin asymmetry dependence of the surface width contribution. 
The given values are the lower and upper limits found for $\Delta S$. }
\label{tab4}
\begin{tabular}{c c c}
%\begin{tabularx}{0.5\textwidth}{c c c}
\hline
\hline
   I  &  &  $\Delta S$ (fm)   \\
\hline    
\hline 
0.02   &  &  0.0032...0.0061    \\              
 0.04   &  &  0.0064...0.012         \\          
 0.06  &  &   0.010...0.018    \\                
 0.08  &  &   0.013...0.024     \\              
 0.1    &  & 0.016...0.030    \\                
 0.12  &  &   0.020...0.036    \\               
 0.14  &  &   0.022...0.043   \\               
 0.16  &  &   0.026...0.049         \\          
 0.18  &  &   0.029...0.055    \\                 
 0.2    &  & 0.032...0.061    \\               
 0.22   &  &  0.035...0.067    \\                  
\hline
\hline
\end{tabular}
\end{table}

\section{Summary and Conclusions }                                                                  
\label{Concl} 

Together with collaborators, recently I derived EoS for NM and SNM based on high-quality 2NFs at N$^3$LO and including all subleading 3NFs. These were used to obtain the EoS of stellar matter and applied in calculations of neutron star radii~\cite{SM22}. In this paper, the focus has been on neutron skins, which are obtained applying the same microscopic EoS in the droplet model. The main intent is to explore average patterns, in particular the relative size of the neutron skin using the same tools across nuclei and predictions.

Conclusions can be summarized as follows. 

 On the theoretical side: nuclear physics has come a long way from the days of one-boson-exchange nucleon-nucleaon potentials and attempts to incorporate some 3NF with no clear scheme or guidance. As for 
 phenomelogical interactions (DFT with mean field models or Skyrme interactions), they are a very useful tool to probe sensitivities and explore correlations, but, by their very nature, cannot address important questions in {\it ab initio} nuclear structure.
Thanks to continuous progress in nuclear theory, one is now able to construct nuclear forces in a systematic and internally consisten manner. The order-by-order structure inherent to chiral EFT allows to explore the importance of different contributions from few-nucleon forces as they emerge at each order. For better understanding of intriguing systems such as neutron skins and neutron stars, it is important to build on that progress.
Predictions from state-of-the-art nuclear theory favor a softer density dependence of the symmetry energy --  on the low-to-medium end of what is considered a realistic range -- and, consistently, smaller values of the neutron skin and the radius of the average-mass neutron star.

 On the experimental side: the symmetry energy parameters that drive the neutron skin are not measured directly, but rather extracted from measurements of suitable observables. While electroweak (EW) methods avoid the uncertainty inherent to the use of hadronic probes, the weakness of the signal seems to generate large errors. This may interfere with the ability of the result to provide a benchmark.
The authors of Ref.~\cite{27Al} state that  ``{\it The EW technique has recently
been applied to $^{208}$Pb and the resulting neutron skin
was found to be in some tension with earlier non-EW
results which favor a thinner skin.The benchmark
of the EW technique which our result can provide
is especially important in light of this observed tension.}" In fact, the tension is better described as irreconcilable differences between essentially all state-of-the-art predictions and the PREX II result. 

\section*{Acknowledgments}
This work was supported by 
the U.S. Department of Energy, Office of Science, Office of Basic Energy Sciences, under Award Number DE-FG02-03ER41270.

\end{document}